\documentclass{aa}

\usepackage{graphicx}
\usepackage{txfonts}
\begin{document}

   \title{Anomalous Z Cam stars: a response to mass--transfer outbursts}

   \author{J.-M. Hameury
          \inst{1}
          \and
          J.-P. Lasota\inst{2,3,4}
          }

   \institute{Observatoire Astronomique de Strasbourg, CNRS UMR 7550, 67000 Strasbourg, France\\
             \email{jean-marie.hameury@astro.unistra.fr}
                 \and
             CNRS, UMR 7095, Institut d'Astrophysique de Paris, 98bis Bd Arago, 75014 Paris, France
              \and
             Nicolaus Copernicus Astronomical Center, Bartycka 18, Warsaw, Poland
         \and
             Sorbonne Universit\'es, UPMC Univ Paris 06, UMR 7095, 98bis Bd Arago, 75014 Paris, France
             }

   \date{}


  \abstract
   {Recent observations of two unusual Z Cam systems, V513 Cas and IW And have shown light curves that seem to contradict the disc-instability model for dwarf novae: outbursts are appearing during standstills of the system when according to the model, the disc is supposed to be in a hot quasi-equilibrium state.}
   {We investigate what additional physical processes need to be included in the model to reconcile it with observations of such anomalous Z Cam systems.}
   {We used our code for modeling thermal-viscous outbursts of the accretion discs and determined what types of mass-transfer variations reproduce the observed light curves.}
   {Outbursts of mass transfer (with a duration of a few days, with a short rise time and an exponential decay) from the stellar companion will account for the observed properties of V513 Cas and IW And, provided they are followed by a short but significant mass-transfer dip. The total mass involved in outbursts is of the order of 10$^{23}$g.}
   {We studied the possible origins of these mass transfer outbursts and showed that they most probably   result from a giant flare near the secondary star surface, possibly due to the absence of star spots in the $L_1$ region.}


   \maketitle
%

   \begin{figure*}
   \centering
   \includegraphics[angle=-90,width=\textwidth]{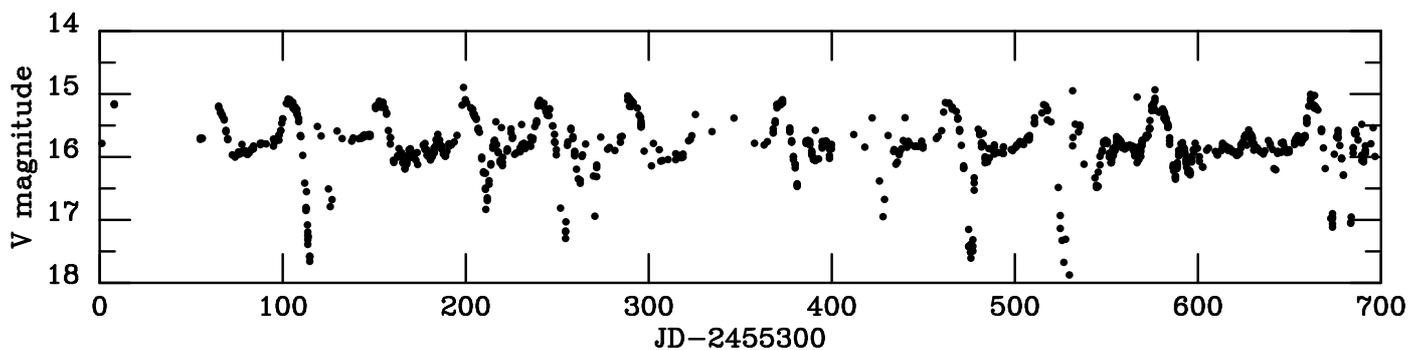}
   \caption{Light curve of V513 Cas (data from AAVSO).}
   \label{lc}%
   \end{figure*}

\section{Introduction}
Dwarf novae are cataclysmic variable stars that undergo recurring outbursts that last several days or more during which the luminosity rises by typically 4-6 magnitudes. These outburts are separated by low-luminosity quiescence intervals that last in general for a few weeks \citep[see e.g.][for a review of observations]{warner95-1}. It is now widely accepted that these outbursts are caused by a thermal/viscous instability of the accretion disc that sets in when the accretion disc is partially ionized \citep[see][for a review of the disc-instability model]{lasota-01}. An accretion disc can only be steady if either its effective temperature is below $\sim 6000$K everywhere, down to the inner disc edge, which means extremely low accretion rates, or if the disc is hot with effective temperatures above 8000K everywhere, up to the outer disc edge, implying accretion rates above a certain critical value $\dot{M}^+_{\rm crit}$. In practice, when the mass transfer from the secondary is lower than $\dot{M}^+_{\rm crit}$, the disc undergoes a cycle of outbursts and quiescent states. An important ingredient of the model is the viscosity, parametrized according to the $\alpha$ prescription of \citet{shakura-sunyaev}; the $\alpha$ parameter must be  ~five times higher in outburst than in quiescence.

Although it suffers from some apparent weaknesses, in particular a simplistic description of viscosity and the the use of 1+1D numerical codes, the disc-instability model has been quite successful in accounting for the basic characteristics of dwarf nova light curves (outburst duration, recurrence time, amplitudes, continuum spectral dependence, etc.). News about its demise based on the alleged (too) large distance to the brightest dwarf nova SS Cyg have been premature \citep{SSCygM} and ultimately false \citep{SSCygR}. Several additional physical ingredients have also been found necessary to account for the variety of outburst types found in nature; most notably, fluctuations of the mass transfer from the secondary (either intrinsic or due to irradiation) and tidal instabilities. Following the observation by \citet{vogt-83} of mass-transfer enhancement during superoutbursts of the SU UMa systems -- a subclass of dwarf novae that have both normal and very long outbursts -- \citet{osaki-85} suggested that irradiation by the accreting white dwarf was responsible for superoutbursts. This author has changed his mind and proposed a tidal instability model \citep{suuma}, in which superoutbursts were due to a tidal instability triggered when the accretion disc expands beyond the 3:1 resonance radius during a normal
outburst. The tidal-thermal instability model was disputed for example by \citet{smak-91}, who noted that observations of \object{Z Cha} and \object{VW Hyi} did appear to support the enhanced mass-transfer hypothesis in SU UMa systems; \citet{smak-waagen} also noted that U Gem underwent a long outburst in 2005, very similar to SU UMa superoutbursts (including superhumps, which are modulations of the light curve with a period slightly different from the orbital period), even though the mass ratio in U Gem is too high for the 3:1 resonance to occur in this system. \citet{cannizzo-12} also noted that \object{U Gem} and \object{SS Cyg} showed embedded precursors in long outbursts, a feature that was thought to be unique of SU UMa systems, and was a specific prediction of the tidal-thermal instability model. It was also shown by \citet{hameury-00} that mass-transfer fluctuations were needed to account for the frantic behaviour of the ER UMa systems, which show sequences of long and short outbursts on short timescales.

Recent observations of several unusual Z Cam systems, \object{IW And}, \object{V513 Cas} \object{HX Peg}, \object{AH Her}, and \object{AT Cnc} by \citet{JAVSO11} brought new evidence that the dwarf nova outburst-cycle behaviour is more complex than initially believed; later observations of IW And and V513 Cas by \citet{szkody-13} confirmed the unusual behaviour of these two systems. Inspection of the AAVSO light curves show that the other three systems appear to be less illustrative. Z Cam systems have mass-transfer rates close to $\dot{M}^+_{\rm crit}$, and exhibit long standstill intervals during which the luminosity is intermediate between quiescence and outbursts. The Z Cam behaviour can be explained by a varying mass-transfer rate:  standstills correspond to periods during which the actual mass-transfer rate is above $\dot{M}^+_{\rm crit}$, and the disc is in a hot, stable state; whereas when the mass transfer slightly decreases below $\dot{M}^+_{\rm crit}$ and enters the instability zone, a sequence of low states and outbursts appears \citep{buat-01}. The model clearly predicts that a standstill can only
be followed by a decline to quiescence, never by an outburst. This is because during standstills the disc is in a hot state in which thermal/viscous outbursts are not possible. This is a very generic prediction of all models in which the system alternates between two stable states. Indeed, until the observations by \citet{JAVSO11}, all Z Cam system standstills were terminated by a decline to quiescence. As can be seen in Fig. \ref{lc}, this is clearly not the case for V513 Cas; outbursts are observed during standstills, with an increase in optical luminosity of slightly less than 1 magnitude; these outbursts are usually followed by a decay to a low state at magnitudes of approximately 18. In some cases, it seems that the disc directly
returns to the standstill level. IW And shows a similar behaviour, but one should note that the data are sparser than for V513 Cas. Recently, the newly identified dwarf nova \object{ST Cha} has been observed to exhibit a light curve similar to that of V513 Cas \citep{JAVSO}.

In Sects. \ref{sec:2} and \ref{sec:3} we show that no physically motivated modifications of the dwarf nova disc-instability model can account for the observed light curves of such anomalous Z Cam stars. In Sect. \ref{sec:4} we show that observations can only be reproduced by an outburst of mass transfer from the secondary, and we then discuss the possible outburst mechanisms. Section \ref{sec:5} contains our conclusions.

\section{Three stable equilibria in a dwarf-nova disc?}
\label{sec:2}

The spectra of IW And observed by \citet{szkody-13} are typical of standard dwarf novae, with strong Balmer lines in emission during quiescence and in absorption during outbursts; the possibility of an abnormal accretion disc, which could have been related for instance to a peculiar chemical composition of the secondary, can therefore be excluded. One cannot, however, totally exclude the possibility that accretion disc equilibria in the three anomalous Z Cam systems form three stable branches, as would be the case if the so-called S-curve that relates the disc effective temperature and the surface density presented a double S-shape.

Indeed, the basis of the disc-instability model is the existence of a thermal-viscous instability in regions where the disc is partially ionized and the opacities are strongly temperature dependent. This instability is conveniently visualized in the surface density $\Sigma$ -- effective temperature $T_{\rm eff}$ plane (or equivalently, the accretion rate $\dot{M}$). At a given radius the equilibrium curve has the well-known S-shape, with cold and hot stable branches, linked by an unstable intermediate segment. Sometimes small wiggles appear on the S-curve \citep[see e.g. Fig. 1 in][]{lasota-01}, which result from convection and changes in molecular opacities, but these are always too small and too localized in radius to generate instabilities; rather, they may cause small undetectable oscillations in the light curve. On the other hand, the shape of the actual S-curve is determined by changing the viscosity parameter $\alpha$ from its quiescence value (typically $\alpha_{\rm c} \simeq 0.04$) to approximately 0.2 on the hot branch. This transition is assumed to occur when partial ionization sets in, but the evidence for this has been circumstantial \citep[see, however,][]{hirose-14}. One could therefore conceive that for some reason, the actual shape of the S-curve is altered and that a new stable branch is generated for example by a more complex change in $\alpha$ than commonly assumed.

The most natural place where to alter the S-curve is the intermediate branch, where partial ionization sets in. This would mean that the systems lies on this new branch during
standstills. This is impossible, however, because this new stable branch would have to fit between the two turning points of the S-curve \citep{LDK}:
\begin{equation}
\dot{M}_{\rm crit}^- = 2.64 ~ 10^{15} ~ \alpha^{-0.04} \left( {M_1 \over
\rm M_\odot} \right)^{-0.89} \left( {r \over 10^{10} \; \rm cm} \right)^{2.67}
~\rm g~s^{-1}
\end{equation}
and
\begin{equation}
\label{eq:mplus}
\dot{M}_{\rm crit}^+ = 8.07 ~ 10^{15} ~ \alpha^{0.01} \left( {M_1 \over
\rm M_\odot} \right)^{-0.85} \left( {r \over 10^{10} \; \rm cm} \right)^{2.58}
~\rm g~s^{-1}
,\end{equation}
where $M_1$ is the primary mass. These turning points are essentially independent of $\alpha$ and have a steep radial dependence; moreover, $\dot{M}_{\rm crit}^-$ and $\dot{M}_{\rm crit}^+$ differ by a factor of about 3. It is thus impossible to find an $\dot{M}$, assumed to be constant throughout the disc during standstills, such that $\dot{M}_{\rm crit}^- < \dot{M} < \dot{M}_{\rm crit}^+$ throughout the disc.

There remains the possibility of altering either the cold or the hot branches. In the first case, standstills would correspond to $\dot{M} < \dot{M}_{\rm crit}^-$ down to the inner disc radius, which would imply accretion rates not compatible with the observed luminosity and spectrum during standstills. Altering the hot branch does not appear to be a solution either, because $\alpha$ is already of the order of 0.1 - 0.2 for classical dwarf novae \citep{KL}, and one would need to reach values close to 1 to create a new branch sufficiently different from the classical one for enabling outbursts.

The above reasoning is based on the assumptions that the changes introduced to create a new branch result from a modification of $\alpha$ alone, and that these modifications are compatible with the standard model for dwarf novae. One would need to severely deviate from these assumptions to be able to reproduce the observed properties of V513 Cas, ST Cha and IW And, which would not make much sense.

\section{Tidal instability of the accretion disc?}
\label{sec:3}

\citet{suuma} proposed that tidal instabilities are responsible for the superoutbursts of the SU UMa systems; these are expected to occur when the disc outer edge reaches the 3:1 resonance radius, which is possible only for systems with low mass ratios ($q < 0.25$). If such an instability were to occur, the maximum possible mass of the secondary would thus be 0.35 M$_\odot$, assuming a primary mass equal to the Chandrasekhar mass. The orbital period of IW And and V513 Cas is 3.7 hr and 5.2 hr; whereas IW And can indeed have such a low mass secondary, this seems most unlikely for V513 Cas, unless the secondary is significantly more oversized than would be expected at this period (see e.g. Fig. 9 in \citet{knigge-11}). ST Cha has a period of 6.84 hr, which places this system safely outside the domain of application of the tidal-thermal instability models.

To be on the safe side we did, however, model a disc evolution assuming the tidal instability sets in at some point. We used our code as described in \citet{buat-02}, artificially increasing both the tidal torque and heating for about ten days, until the disc had shrunk enough for the tidal instability to stop, which
restored the tidal term to its normal value. The average mass transfer rate was kept constant and slightly higher than $\dot{M}_{\rm crit}^+ $. Outbursts were produced, as expected, but the system always returned to standstill at the end of an outburst instead of entering quiescence. This last feature could be reproduced only if the mass transfer from the secondary were reduced at the same time (or shortly afterwards) as the tidal instability sets in. This could occur if the expansion of the disc were precisely due to a reduction of mass transfer during standstill, provided that this reduction were gradual enough so that a cooling wave did not start immediately from the outer edge, stopping the disc expansion, and lasted for about the time it takes for the disc to have contracted beyond the point where the tidal instability stops. All of this requires a fine-tuning of parameters that are not related to any physical phenomenon, which makes this
scenario very unlikely.

We conclude therefore that the tidal instability cannot account for the light curves of anomalous Z~Cam stars. This is consistent with the fact that this instability is not expected to occur in systems under consideration here.
\begin{figure}
   \centering
   \includegraphics[width=\columnwidth]{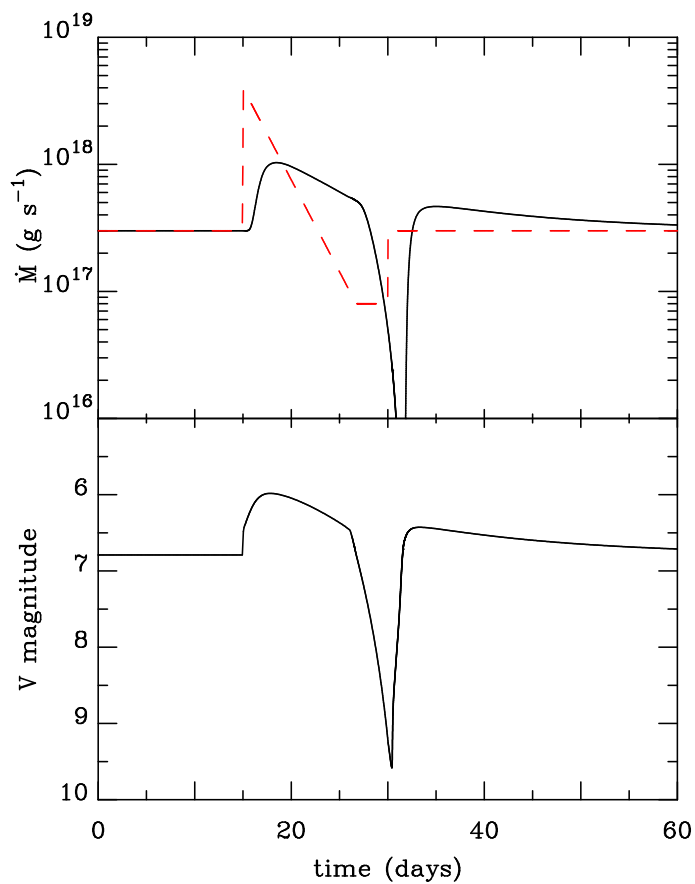}
   \caption{Response of the accretion disc to a mass-transfer outburst. The top panel shows the mass-transfer rate from the secondary (dashed red curve) and the mass-accretion rate onto the white dwarf (solid curve), the bottom panel shows the visual magnitude.}
   \label{model}%
   \end{figure}

\section{Outburst of mass transfer}
\label{sec:4}

We now examine the only possibility left: outbursts in the three anomalous Z Cam stars are due to a mass-transfer outburst from the secondary.

\subsection{Disc response to mass-transfer outbursts}

We have modelled the disc response to a mass transfer outburst using our code described in \citet{buat-02}, assuming a primary mass of 0.8 M$_\odot$, a secondary mass of 0.5 M$_\odot,$ and an orbital period of 5.2 hr, corresponding to V513 Cas. The average outer disc radius is then $4.295\times 10^{10}$cm. The viscosity parameter $\alpha$ was assumed to be $\alpha_{\rm c} = 0.04$ on the cold and $\alpha_{\rm h} = 0.2$ on the hot branch, with the transition from the cold to hot situation described as in \citet{hameuryetal98-1}. We found that the anomalous Z Cam behaviour is reproduced by assuming the time profile of the mass-transfer rate from the secondary shown in Fig. \ref{model}; we started from a steady-state situation corresponding to standstills where the mass transfer rate is $3 \times 10^{17}$ g~s$^{-1}$ (slightly above the critical accretion rate at the outer disc rim\footnote{This value does not correspond to Eq. (\ref{eq:mplus}) at the assumed outer radius because in the code additional heating by e.g. tidal torques is taken into account. Such effects lower the value of the critical accretion rate \citep[see][]{buat-01}.} $2.3\times 10^{17} g~s^{-1}$), which was then abruptly increased to $4 \times 10^{18}$ g~s$^{-1}$ and then decreased exponentially to a minimum of $8 \times 10^{16}$ g~s$^{-1}$ before recovering the quasi-steady-state value.

As can be seen, the time evolution of the optical light curve is quite similar to that observed in V513 Cas; the outburst amplitude is of the order of 0.8 mag, with a rapid rise (about a day) and a slightly decaying plateau, followed by a rapid decline to a short quiescent phase. The rise time in the optical is controlled by $\alpha_{\rm h}$, provided that the mass transfer rises on a shorter timescale, and the outburst duration is approximately equal to the duration of the mass-transfer outburst. The existence of a quiescent state is determined by the fact that the mass transfer from the secondary decreases below $\dot{M}_{\rm crit}^{+}$ at the end of an outburst for a time long enough for a cooling wave to start from the outer disc edge; as soon as the the mass-transfer rate is restored to its standstill value, an outside-in outburst begins, possibly before the cooling wave has had enough time to reach the inner disc edge (this is the situation depicted in Fig. \ref{model}). As the propagation time of the heating front is much shorter than that of a cooling front, the disc may never have reached a full quiescence state. This explains both the short duration of the quiescent states shown in Fig. \ref{lc} and their varying depth, and the fact that in some cases the system returns directly to standstill at the end of an outburst (e.g. at JD 2455600).

We can thus account for the observed light curves, at the expense of rather strong requirements on the shape of mass transfer outbursts. The most stringent constraints are (1) the large amplitude of the outburst, and (2) the requirement that the mass-transfer rate decreases below $\dot{M}_{\rm crit}^{+}$ at the end of an outburst before returning to its steady-state value. On the other hand, it could be argued that the light curves of the three anomalous Z Cam systems are a strong argument in favour of the existence of such outbursts since no other plausible explanation of their shape seems to exist.

\subsection {Origin of the mass-transfer outbursts}

The important characteristics of the mass-transfer outbursts needed to account for the observed light curves are the total mass involved during the outbursts and their relative regularity,  in terms of recurrence time, amplitude, and durations. However, these outbursts are not found in most systems and may thus be due to some rare set of circumstances/parameters.

\subsubsection{Coronal mass ejections}

The total amount of mass transferred during the outburst shown in Fig \ref{model} in is $6.5 \times 10^{23}$ g, orders of magnitude above what is observed in solar coronal mass ejections (CMEs), which is on average of the order of 10$^{15}$ g and can reach at most about 10$^{17}$g; their duration is also much shorter \citep{cme} than that of anomalous Z Cam outbursts. We do not know much about CMEs in other stars, in particular those of the lower main sequence, which are much more active than the Sun. It is unlikely, however, that the mass involved in stellar CMEs is large enough to account for the outbursts observed in V515 Cas and IW And.

\subsubsection{Magnetic activity}
Magnetic activity has been invoked to account for the low states of AM Her systems, which do not have accretion discs, and whose luminosity changes can thus be directly attributed to changes in mass-transfer rates. \citet{livio-pringle-94} and \citet{king-cannizzo-98} argued that large star spots entering the $L_1$ area would strongly reduce the mass-transfer rate, but this proposal was challenged by \citet{howell-00}. Although star spots have been put forward to account for low states, they could also produce outbursts in a situation where they are so numerous that on the average, at least one star spot would be expected in the $L_1$ area; outbursts could then be produced when there is no star spot close to $L_1$. Low states in AM Her systems last much longer than the outbursts  in V513 Cas, indicating that star spots move or are generated on long timescales in these systems. The low states of the AM Her star cannot be attributed to their full convective structure (as might be suggested by the periods of most of them, which are shorter than 2 hr)  because such systems at periods > 3 hr also display low states. The very presence of AM Her stars at long periods could be due to long-term mass-transfer fluctuations \citep{HKL89}. In addition, VY Scl stars whose secondary stars also stop transferring matter to the primary white dwarf all have orbital periods of between three and four hours \citep{warner95-1} and are assumed to be still in possession of their radiative cores. Clearly, the universal phenomenon of mass-transfer variations in CVs is not understood at all.

\subsubsection{Irradiation of the secondary}
As mentioned earlier in this paper, irradiation has been proposed as a possible mechanism for the outbursts of dwarf novae and to account for the high and low states of AM Her systems \citep{king-lasota-84}. Irradiation was also invoked \citep{hameury-00} to account for the long duration of outbursts in systems such as \object{RZ LMi} or \object{EG Cnc}. There is direct evidence for modulated irradiation of the secondary stars during dwarf nova superoutbursts \citep{smak-11}.  It is not clear, however, that irradiation could be sufficient to account for the observed outburst shape and intensity; except for AM Her systems --- which do show large mass-transfer variations, but not, strictly speaking, mass-transfer outbursts --- (but there the secondary is irradiated by hard X-rays, which can penetrate deep below the photosphere), the variations in the mass-transfer rate are assumed to be relatively small. It is not clear either what would set V513 Cas, ST Cha and IW And apart from other systems.

\subsubsection{A third body}
The existence of a third body, be it a planet such as the one found in \object{V893 Sco} \citep{bruch-14} or a more massive companion, could account for periodic or quasi-periodic variations of the mass-transfer rate by modulating the orbital separation. \citet{chavez-12} suggested that this could be the explanation for the long-term modulation observed in the light curve of \object{FS Aur}. Although the outburst sequence is very regular (see Fig. \ref{lc}), it is not strictly periodic; moreover, in such a scenario it would be extremely difficult to reproduce the very short rise time of the mass-transfer outburst.

\section{Conclusions}
\label{sec:5}
The disc-instability model can be reconciled with the observations of  the anomalous Z Cam stars V513 Cas, ST Cha, and IW And, which exhibit outbursts during standstills provided that these outbursts result from outbursts in mass transfer from the secondary. The observed light curves are very well reproduced, but the reason for the required mass-transfer outbursts remains rather elusive. The most promising explanation is that they are due to  magnetic activity of the secondary star in these three systems, which might be much stronger than in other systems, thus explaining why they do not behave as all other Z Cam systems, in which the standstills are interrupted by a decline to a low state, with outbursts starting only during low states.

\begin{acknowledgements}
We acknowledge with thanks the variable-star observations from the AAVSO International Database contributed by observers worldwide that were used in this research.
JPL acknowledges support from the French Space Agency CNES and Polish NCN grants UMO-2011/01/B/ST9/05439 and UMO-2013/08/A/ST9/00795.
\end{acknowledgements}


\end{document}